\documentstyle[twocolumn,aps,epsfig,floats]{revtex}

\begin{document}

\def\eref#1{(\ref{#1})}
\def\vv{{\bf v}}
\def\vM{{\bf M}}
\def\O{{\rm O}}

\twocolumn[\hsize\textwidth\columnwidth\hsize\csname @twocolumnfalse\endcsname

\title{Epidemics and percolation in small-world networks}
\author{Cristopher Moore$^{1,2}$ and M. E. J. Newman$^1$}
\address{$^1$Santa Fe Institute, 1399 Hyde Park Road, Santa Fe,
New Mexico 87501}
\address{$^2$Departments of Computer Science and Physics, University of New
Mexico, Albuquerque, New Mexico 87131}
\maketitle

\begin{abstract}  
  We study some simple models of disease transmission on small-world
  networks, in which either the probability of infection by a disease or
  the probability of its transmission is varied, or both.  The resulting
  models display epidemic behavior when the infection or transmission
  probability rises above the threshold for site or bond percolation on the
  network, and we give exact solutions for the position of this threshold
  in a variety of cases.  We confirm our analytic results by numerical
  simulation.
\end{abstract}

\pacs{87.23.Ge, 05.10.Cc, 05.70.Jk, 64.60.Fr}

]

\section{Introduction}
It has long been recognized that the structure of social networks plays an
important role in the dynamics of disease propagation.  Networks showing
the ``small-world'' effect~\cite{Kocken89,Watts99}, where the number of
``degrees of separation'' between any two members of a given population is
small by comparison with the size of the population itself, show much
faster disease propagation than, for instance, simple diffusion models on
regular lattices.

Milgram~\cite{Milgram67} was one of the first to point out the existence of
small-world effects in real populations.  He performed experiments which
suggested that there are only about six intermediate acquaintances
separating any two people on the planet, which in turn suggests that a
highly infectious disease could spread to all six billion people on the
planet in only about six incubation periods of the disease.

Early models of this phenomenon were based on random
graphs~\cite{SS88,KM96}.  However, random graphs lack some of the crucial
properties of real social networks.  In particular, social networks show
``clustering,'' in which the probability of two people knowing one another
is greatly increased if they have a common acquaintance~\cite{WS98}.  In a
random graph, by contrast, the probability of there being a connection
between any two people is uniform, regardless of which two you choose.

Watts and Strogatz~\cite{WS98} have recently suggested a new ``small-world
model'' which has this clustering property, and has only a small average
number of degrees of separation between any two individuals.  In this paper
we use a variant of the Watts--Strogatz model~\cite{NW99} to investigate
disease propagation.  In this variant, the population lives on a
low-dimensional lattice (usually a one-dimensional one) where each site is
connected to a small number of neighboring sites.  A low density of
``shortcuts'' is then added between randomly chosen pairs of sites,
producing much shorter typical separations, while preserving the clustering
property of the regular lattice.

Newman and Watts~\cite{NW99} gave a simple differential equation model for
disease propagation on an infinite small-world graph in which communication
of the disease takes place with 100\% efficiency---all acquaintances of an
infected person become infected at the next time-step.  They solved this
model for the one-dimensional small-world graph, and the solution was later
generalized to higher dimensions~\cite{Moukarzel99} and to finite-sized
lattices~\cite{NMW99}.  Infection with 100\% efficiency is not a
particularly realistic model however, even for spectacularly virulent
diseases like Ebola fever, so Newman and Watts also suggested using a site
percolation model for disease spreading in which some fraction $p$ of the
population are considered susceptible to the disease, and an initial
outbreak can spread only as far as the limits of the connected cluster of
susceptible individuals in which it first strikes.  An epidemic can occur
if the system is at or above its percolation threshold where the size of
the largest cluster becomes comparable with the size of the entire
population.  Newman and Watts gave an approximate solution for the fraction
$p_c$ of susceptible individuals at this epidemic point, as a function of
the density of shortcuts on the lattice.  In this paper we derive an exact
solution, and also look at the case in which transmission between
individuals takes place with less than 100\% efficiency, which can be
modeled as a bond percolation process.

\section{Site percolation}
Two simple parameters of interest in epidemiology are {\em susceptibility},
the probability that an individual exposed to a disease will contract it,
and {\em transmissibility}, the probability that contact between an
infected individual and a healthy but susceptible one will result in the
latter contracting the disease.  In this paper, we assume that a disease
begins with a single infected individual.  Individuals are represented by
the sites of a small-world model and the disease spreads along the bonds,
which represent contacts between individuals.  We denote the sites as being
occupied or not depending on whether an individual is susceptible to the
disease, and the bonds as being occupied or not depending on whether a
contact will transmit the disease to a susceptible individual.  If the
distribution of occupied sites or bonds is random, then the problem of when
an epidemic takes place becomes equivalent to a standard percolation
problem on the small-world graph: what fraction $p_c$ of sites or bonds
must be occupied before a ``giant component'' of connected sites forms
whose size scales extensively with the total number $L$ of
sites~\cite{note1}?

We will start with the site percolation case, in which every contact of a
healthy but susceptible person with an infected person results in
transmission, but less than 100\% of the individuals are susceptible.  The
fraction $p$ of occupied sites is precisely the susceptibility defined
above.

\begin{figure}[t]
\begin{center}
\psfig{figure=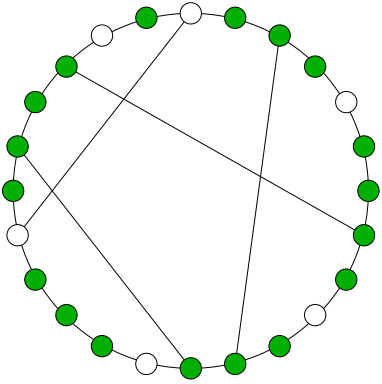,width=2in}
\end{center}
\caption{A small-world graph with $L=24$, $k=1$, and four shortcuts.  The
  colored sites represented susceptible individuals.  The susceptibility is
  $p=\frac34$ in this example.}
\label{site}
\end{figure}

Consider a one-dimensional small-world graph as in Fig.~\ref{site}.
$L$~sites are arranged on a one-dimensional lattice with periodic boundary
conditions and bonds connect all pairs of sites which are separated by a
distance of $k$ or less.  (For simplicity we have chosen $k=1$ in the
figure.)  Shortcuts are now added between randomly chosen pairs of sites.
It is standard to define the parameter $\phi$ to be the average number of
shortcuts per bond on the underlying lattice.  The probability that two
randomly chosen sites have a shortcut between them is then
\begin{equation}
\psi = 1 - \biggl[1-{2\over L^2}\biggr]^{k \phi L} \simeq {2k \phi \over L}
\end{equation}
for large $L$.

A connected cluster on the small-world graph consists of a number of {\em
  local clusters}---occupied sites which are connected together by the
near-neighbor bonds on the underlying one-dimensional lattice---which are
themselves connected together by shortcuts.  For $k=1$, the average number
of local clusters of length $i$ is
\begin{equation}
N_i = (1-p)^2 p^i L.
\end{equation}
For general $k$ we have
\begin{eqnarray}
N_i &=& (1-p)^{2k} p (1-(1-p)^k)^{i-1} L\nonumber\\
    &=& (1-q)^2 p q^{i-1} L,
\end{eqnarray}
where $q = 1-(1-p)^k$.

Now we build a connected cluster out of these local clusters as follows.
We start with one particular local cluster, and first add to it all other
local clusters which can be reached by traveling along a single shortcut.
Then we add all local clusters which can be reached from those new ones by
traveling along a single shortcut, and so forth until the connected cluster
is complete.  Let us define a vector $\vv$ at each step in this process,
whose components $v_i$ are equal to the probability that a local cluster of
size $i$ has just been added to the overall connected cluster.  We wish to
calculate the value $\vv'$ of this vector in terms of its value $\vv$ at
the previous step.  At or below the percolation threshold the $v_i$ are
small and so the $v_i'$ will depend linearly on the $v_i$ according to a
transition matrix $\vM$ thus:
\begin{equation}
v_i' = \sum_j M_{ij} v_j,
\label{itereqn}
\end{equation}
where
\begin{equation}
M_{ij} = N_i (1 - (1-\psi)^{ij}).
\end{equation}
Here $N_i$ is the number of local clusters of size $i$ as before, and
$1-(1-\psi)^{ij}$ is the probability of a shortcut from a local cluster of
size $i$ to one of size $j$, since there are $ij$ possible pairs of sites
by which these can be connected.  Note that $M$ is not a stochastic matrix,
i.e.,~it is not normalized so that its rows sum to unity.

Now consider the largest eigenvalue $\lambda$ of $\vM$.  If $\lambda<1$,
iterating Eq.~\eref{itereqn} makes $\vv$ tend to zero, so that the rate at
which new local clusters are added falls off exponentially, and the
connected clusters are finite with an exponential size distribution.
Conversely, if $\lambda>1$, $\vv$ grows until the size of the cluster
becomes limited by the size of the whole system.  The percolation threshold
occurs at the point $\lambda=1$.

For finite $L$ finding the largest eigenvalue of $M$ is difficult.
However, if $\phi$ is held constant, $\psi$ tends to zero as $L\to\infty$,
so for large $L$ we can approximate $\vM$ as
\begin{equation}
M_{ij} = ij \psi N_i.
\end{equation}
This matrix is the outer product of two vectors, with the result that
Eq.~\eref{itereqn} can be simplified to
\begin{equation}
\lambda v_i = i \psi N_i \sum_j j v_j,
\end{equation}
where we have set $v'_i = \lambda v_i$.  Thus the eigenvectors of $\vM$
have the form $v_i = C\lambda^{-1} i\psi N_i$ where $C = \sum_j j v_j$ is a
constant.  Eliminating $C$ then gives
\begin{equation}
\lambda = \psi \sum_j j^2 N_j.
\label{lambdaeqn}
\end{equation}
For $k=1$, this gives
\begin{equation}
\lambda = \psi L p \, \frac{1+p}{1-p} = 2 \phi p \, \frac{1+p}{1-p}.
\label{sitelambdak1}
\end{equation}
Setting $\lambda=1$ yields the value of $\phi$ at the percolation threshold
$p_c$:
\begin{equation}
\phi = \frac{1-p_c}{2 p_c \, (1+p_c)},
\end{equation}
and solving for $p_c$ gives
\begin{equation}
p_c = \frac{\sqrt{4 \phi^2 + 12 \phi + 1} - 2 \phi - 1}{4 \phi}.
\label{sitethreshk1}
\end{equation}
For general $k$, we have
\begin{equation}
\lambda = \psi L p \, \frac{1+q}{1-q} 
= 2k \phi p \, \frac{2-(1-p)^k}{(1-p)^k},
\end{equation}
or, at the threshold
\begin{equation}
\phi = \frac{(1-p_c)^k}{2 k p_c \, (2-(1-p_c)^k)}.
\label{sitethreshgenk}
\end{equation}
The threshold density $p_c$ is then a root of a polynomial of order $k+1$.

\section{Bond percolation}
An alternative model of disease transmission is one in which all
individuals are susceptible, but transmission takes place with less than
100\% efficiency.  This is equivalent to bond percolation on a small-world
graph---an epidemic sets in when a sufficient fraction $p_c$ of the {\em
bonds\/} on the graph are occupied to cause the formation of a giant
component whose size scales extensively with the size of the graph.  In
this model the fraction $p$ of occupied bonds is the transmissibility of
the disease.

For $k=1$, the percolation threshold $p_c$ for bond percolation is the same
as for site percolation for the following reason.  On the one hand, a local
cluster of $i$ sites now consists of $i-1$ occupied bonds with two
unoccupied ones at either end, so that the number of local clusters of $i$
sites is
\begin{equation}
N_i = (1-p)^2 \, p^{i-1},
\end{equation}
which has one less factor of $p$ than in the site percolation case.  On the
other hand, the probability of a shortcut between two random sites now has
an extra factor of $p$ in it---it is equal to $\psi p$ instead of just
$\psi$.  The two factors of $p$ cancel and we end up with the same
expression for the eigenvalue of $\vM$ as before, Eq.~\eref{sitelambdak1},
and the same threshold density, Eq.~\eref{sitethreshk1}.

For $k>1$, calculating $N_i$ is considerably more difficult.  We will solve
the case $k=2$.  Let $Q_i$ denote the probability that a given site $n$ and
the site $n-1$ to its left are part of the same local cluster of size $i$,
when only bonds to the left of site $n$ are taken into account.  Similarly,
let $Q_{ij}$ be the probability that $n$ and $n-1$ are part of two separate
local clusters of size $i$ and $j$ respectively, again when only bonds to
the left of $n$ are considered.  Then, by considering site $n+1$ which has
possible connections to both $n$ and $n-1$, we can show that
\begin{equation}
Q_{i+1,j} = \left\lbrace \begin{array}{ll}
            (1-p)^2 \bigl[ Q_j + \sum_k Q_{jk} \bigr] \quad &
            \mbox{for $i=0$}\\
            p(1-p) \,Q_{ji} & \mbox{for $i\ge1$,}
            \end{array}\right.
\end{equation}
and
\begin{eqnarray}
Q_{i+1} &=& p(2-p) \,Q_i + p(1-p) \sum_j Q_{ij} +
            p^2 \!\! \sum_{j+j'=i} \!\! Q_{jj'}.\nonumber\\
\end{eqnarray}
If we define generating functions $H(z) = \sum_i Q_i z^i$ and $H(z,w) =
\sum_{i,j} Q_{ij} z^i w^j$, this gives us
\begin{eqnarray}
\label{heqn1}
H(z,w) &=& z(1-p)^2 \bigl[ H(w) + H(w,1) \bigr]\nonumber\\
       & & \quad +\, zp(1-p) \,H(w,z),\\
\label{heqn2}
H(z)   &=& zp(2-p) H(z) + zp(1-p) H(z,1)\nonumber\\
       & & \quad +\, zp^2 H(z,z).
\end{eqnarray}
Since any pair of adjacent sites must belong to some cluster or clusters
(possibly of size one), the probabilities $Q_i$ and $Q_{ij}$ must sum to
unity according to $\sum_i Q_i + \sum_{i,j} Q_{ij} = 1$, or equivalently
$H(1) + H(1,1) = 1$.  Finally, the density of clusters of size $i$ is equal
to the probability that a randomly chosen site is the rightmost site of
such a cluster, in which case neither of the two bonds to its right are
occupied.  Taken together, these results imply that the generating function
for clusters, $G(z) = \sum_i N_i z^i$, must satisfy
\begin{equation} 
G(z) = (1-p)^2 \bigl[ H(z) + H(z,1) \bigr].
\end{equation}
Solving Eqs.~\eref{heqn1} and~\eref{heqn2} for $H(z)$ and $H(z,1)$ then
gives
\begin{eqnarray}
G(z) &=& \bigl[z (1 - p)^4 \bigl(1 - 2 p z + p^3 (1 - z) z + p^2 z^2
         \bigr)\bigr]\big/\nonumber\\
& & \quad \bigl[1 - 4 p z + p^5 (2 - 3z) z^2 - p^6 (1 - z) z^2\nonumber\\
& & \qquad +\, p^4 z^2 (1 + 3z) + p^2 z (4 + 3z)\nonumber\\
& & \qquad -\, p^3 z \bigl(1 + 5z + z^2\bigr)\bigr],
\end{eqnarray}
the first few terms of which give
\begin{eqnarray}
N_1/L &=& (1-p)^4,\\
N_2/L &=& 2 p \, (1-p)^6,\\
N_3/L &=& p^2 \, (1-p)^6 \, (6 - 8 p + 3 p^2).
\end{eqnarray}
Again replacing $\psi$ with $\psi p$, Eq.~\eref{lambdaeqn} becomes
\begin{equation}
\lambda = \psi p \sum_i i^2 N_i = 2k \phi p \biggl[ \biggl( z \frac{d}{dz}
  \biggr)^{\!2} G(z) \,\biggr]_{z=1},
\label{bondlambda}
\end{equation}
which, setting $k=2$, implies that the percolation threshold $p_c$ is given
by
\begin{equation}
\phi = \frac{(1 - p_c)^3 \, (1 - p_c + p_c^2)} 
       {4 p_c \,(1 + 3 p_c^2 - 3 p_c^3 - 2 p_c^4 + 5 p_c^5 - 2 p_c^6)}.
\label{bondthreshk2}
\end{equation}
In theory it is possible to extend the same method to larger values of $k$,
but the calculation rapidly becomes tedious and so we will, for the moment
at least, move on to other questions.

\section{Site and bond percolation}
The most general disease propagation model of this type is one in which
both the susceptibility and the transmissibility take arbitrary values,
i.e.,~the case in which sites and bonds are occupied with probabilities
$p_{\rm site}$ and $p_{\rm bond}$ respectively.  For $k=1$, a local cluster
of size $i$ then consists of $i$ susceptible individuals with $i-1$
occupied bonds between them, so that
\begin{equation}
N_i = (1 - p_{\rm site} \,p_{\rm bond})^2 
                    \,p_{\rm site}^i \,p_{\rm bond}^{i-1}.
\end{equation}
Replacing $\psi$ with $\psi p_{\rm bond}$ in Eq.~\eref{lambdaeqn} gives
\begin{equation}
\lambda = \psi \,p_{\rm bond} \sum_j j^2 N_j = 2 \phi p \,\frac{1+p}{1-p},
\end{equation}
where $p = p_{\rm site} \,p_{\rm bond}$.  In other words, the position of
the percolation transition is given by precisely the same expression as
before except that $p$ is now the product of the site and bond
probabilities.  The critical value of this product is then given by
Eq.~\eref{sitethreshk1}.  The case of $k>1$ we leave as an open problem for
the interested (and courageous) reader.

\section{Numerical calculations}
\begin{figure}[t]
\begin{center}
\psfig{figure=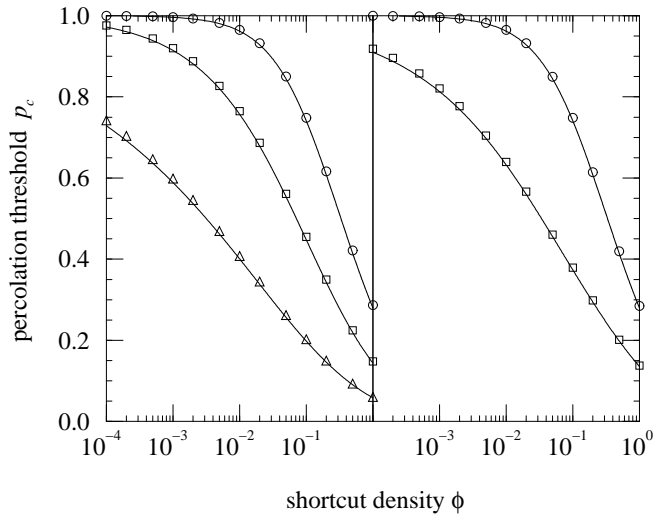,width=\columnwidth}
\end{center}
\caption{The points are numerical results for the percolation threshold as
a function of shortcut density~$\phi$ for systems of size $L=10^6$.  Left
panel: site percolation with $k=1$ (circles), $2$~(squares), and
$5$~(triangles).  Right panel: bond percolation with $k=1$~(circles) and
$2$~(squares).  Each point is averaged over 1000 runs and the resulting
statistical errors are smaller than the points.  The solid lines are the
analytic expressions for the same quantities, Eqs.~\eref{sitethreshk1},
\eref{sitethreshgenk}, and~\eref{bondthreshk2}.  The slight systematic
tendency of the numerical results to overestimate the value of $p_c$ is a
finite-size effect which decreases both with increasing system size and
with increasing $\phi$\protect\cite{note2}.}
\label{perc}
\end{figure}

We have performed extensive computer simulations of percolation and disease
spreading on small-world networks, both as a check on our analytic results,
and to investigate further the behavior of the models.  First, we have
measured the position of the percolation threshold for both site and bond
percolation for comparison with our analytic results.  A naive algorithm
for doing this fills in either the sites or the bonds of the lattice one by
one in some random order and calculates at each step the size of either the
average or the largest cluster of connected sites.  The position of the
percolation threshold can then be estimated from the point at which the
derivative of this size with respect to the number of occupied sites or
bonds takes its maximum value.  Since there are $\O(L)$ sites or bonds on
the network in total and finding the clusters takes time $\O(L)$, such an
algorithm runs in time $\O(L^2)$.  A more efficient way to perform the
calculation is, rather than recreating all clusters at each step in the
algorithm, to calculate the new clusters from the ones at the previous
step.  By using a tree-like data structure to store the
clusters~\cite{BN99}, one can in this way reduce the time needed to find
the value of $p_c$ to $\O(L\log L)$.  In Fig.~\ref{perc} we show numerical
results for $p_c$ from calculations of the largest cluster size using this
method for systems of one million sites with various values of $k$, for
both bond and site percolation.  As the figure shows, the results agree
well with our analytic expressions for the same quantities over several
orders of magnitude in $\phi$.

\begin{figure}[t]
\begin{center}
\psfig{figure=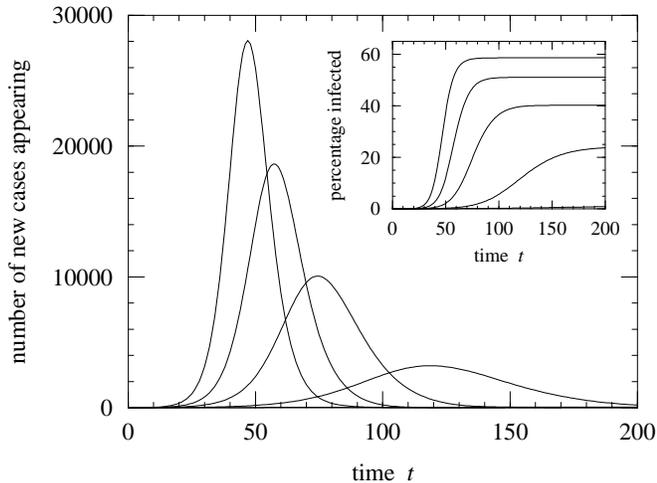,width=\columnwidth}
\end{center}
\caption{The number of new cases of a disease appearing as a function of
  time in simulations of the site-percolation model with $L=10^6$, $k=5$,
  and $\phi=0.01$.  The top four curves are for $p=0.60$, $0.55$, $0.50$,
  and~$0.45$, all of which are above the predicted percolation threshold of
  $p_c=0.401$ and show evidence of the occurrence of substantial epidemics.
  A fifth curve, for $p=0.40$, is plotted but is virtually invisible next
  to the horizontal axis because even fractionally below the percolation
  threshold no epidemic behavior takes place.  Each curve is averaged over
  1000 repetitions of the simulation.  Inset: the total percentage of the
  population infected as a function of time in the same simulations.}
\label{invasion}
\end{figure}

Direct confirmation that the percolation point in these models does indeed
correspond to the threshold at which epidemics first appear can also be
obtained by numerical simulation.  In Fig.~\ref{invasion} we show results
for the number of new cases of a disease appearing as a function of time in
simulations of the site-percolation model.  (Very similar results are found
in simulations of the bond-percolation model.)  In these simulations we
took $k=5$ and a value of $\phi=0.01$ for the shortcut density, which
implies, following Eq.~\eref{sitethreshgenk}, that epidemics should appear
if the susceptibility within the population exceeds $p_c=0.401$.  The
curves in the figure are (from the bottom upwards) $p=0.40$, $0.45$,
$0.50$, $0.55$, and~$0.60$, and, as we can see, the number of new cases of
the disease for $p=0.40$ shows only a small peak of activity (barely
visible along the lower axis of the graph) before the disease fizzles out.
Once we get above the percolation threshold (the upper four curves) a large
number of cases appear, as expected, indicating the onset of epidemic
behavior.  In the simulations depicted, epidemic disease outbreaks
typically affected between 50\% and 100\% of the susceptible individuals,
compared with about 5\% in the sub-critical case.  There is also a
significant tendency for epidemics to spread more quickly (and in the case
of self-limiting diseases presumably also to die out sooner) in populations
which have a higher susceptibility to the disease.  This arises because in
more susceptible populations there are more paths for the infection to take
from an infected individual to an uninfected one.  The amount of time an
epidemic takes to spread throughout the population is given by the average
radius of (i.e.,~path length within) connected clusters of susceptible
individuals, a quantity which has been studied in Ref.~\onlinecite{NW99}.

In the inset of Fig.~\ref{invasion} we show the overall (i.e.,~integrated)
percentage of the population affected by the disease as a function of time
in the same simulations.  As the figure shows, this quantity takes a
sigmoidal form similar to that seen also in random-graph
models~\cite{SS88,KM96}, simple small-world disease models~\cite{NW99}, and
indeed in real-world data.

\section{Conclusions}
We have derived exact analytic expressions for the percolation threshold on
one-dimensional small-world graphs under both site and bond percolation.
These results provide simple models for the onset of epidemic behavior in
diseases for which either the susceptibility or the transmissibility is
less than 100\%.  We have also looked briefly at the case of simultaneous
site and bond percolation, in which both susceptibility and
transmissibility can take arbitrary values.  We have performed extensive
numerical simulations of disease outbreaks in these models, confirming both
the position of the percolation threshold and the fact that epidemics take
place above this threshold only.

Finally, we point out that the method used here can in principle give an
exact result for the site or bond percolation threshold on a small-world
graph with any underlying lattice for which we can calculate the density of
local clusters as a function of their size.  If, for instance, one could
enumerate lattice animals on lattices of two or more dimensions, then the
exact percolation threshold for the corresponding small-world model would
follow immediately.

\section*{Acknowledgments}
We thank Duncan Watts for helpful conversations.  This work was supported
in part by the Santa Fe Institute and DARPA under grant number ONR
N00014-95-1-0975.


\begin{thebibliography}{99}
\bibitem{note1}
Technically this point is not a percolation transition, since it is not
possible to create a system-spanning (i.e.,~percolating) cluster on a
small-world graph.  To be strictly correct, we should refer to the
transition as ``the point at which a giant component first forms.''  This
is something of a mouthful however, so we will bend the rules a little and
continue to talk about ``percolation'' in this paper.
\bibitem{note2}
We have also experimented with finite-size scaling as a method for
calculating $p_c$.  Such calculations are however prone to inaccuracy
because, as pointed in Ref.~\onlinecite{NW99}, the correlation length
scaling exponent $\nu$, which governs the behavior of $p_c$ as system size
is varied, depends on the effective dimension of the graph, which itself
changes with system size.  Thus the exponent can at best only be considered
approximately constant in a finite-size scaling calculation.  Rather than
introduce unknown errors into the value of $p_c$ as a result of this
approximation, we have opted in the present work to quote direct
measurements of $p_c$ on large systems as our best estimate of the
percolation threshold.
\end{thebibliography}
\end{document}